\documentclass[5p,times]{elsarticle}

\usepackage{hyperref}

\journal{Computers in Biology and Medicine}
\begin{document}

\begin{frontmatter}

\title{Mass Segmentation in Automated 3-D Breast Ultrasound Using Dual-Path U-net}

\author{Hamed Fayyaz\corref{mycorrespondingauthor}}
\address{Department of Computer and Information Sciences, University of Delaware, USA}

\cortext[mycorrespondingauthor]{Corresponding author, Tel.:+13026901905, Email: fayyaz@udel.edu}

\author{Ehsan Kozegar}
\address{Department of Computer Engineering, Faculty of Technology and Engineering, University of Guilan, Rudsar-Vajargah, Iran}

\author{Tao Tan}
\address{Department of Mathematics and Computer Science, Eindhoven University of Technology, Eindhoven, The Netherlands}

\author{Mohsen Soryani}
\address{School of Computer Engineering, Iran University of Science and Technology, Tehran, Iran}

\begin{abstract}
		Automated 3-D breast ultrasound (ABUS) is a newfound system for breast screening that has been proposed as a supplementary modality to mammography for breast cancer detection. While ABUS has better performance in dense breasts, reading ABUS images is exhausting and time-consuming. So, a computer-aided detection system is necessary for interpretation of these images. Mass segmentation plays a vital role in the computer-aided detection systems and it affects the overall performance. Mass segmentation is a challenging task because of the large variety in size, shape, and texture of masses. Moreover, an imbalanced dataset makes segmentation harder. A novel mass segmentation approach based on deep learning is introduced in this paper. The deep network that is used in this study for image segmentation is inspired by U-net, which has been used broadly for dense segmentation in recent years. The system's performance was determined using a dataset of 50 masses including 38 malign and 12 benign lesions. The proposed segmentation method attained a mean Dice of 0.82 which outperformed a two-stage supervised edge-based method with a mean Dice of 0.74 and an adaptive region growing method with a mean Dice of 0.65.
\end{abstract}


\end{frontmatter}

\section{Introduction}
Breast cancer is the most common diagnosed cancer and the second leading cause of cancer death among women\cite{siegel2018cancer}. Early detection of metastatic diseases like cancer reduces the mortality rate and breast screening is the best way of lesion detection at early stages. Currently, mammography is the primary modality for screening, which yield a poor sensitivity when breast density is high\cite{boyd2007mammographic}. Besides, exposure to X-ray may increase the risk of incidence of breast cancer for pregnant women and women under 30\cite{pijpe2012exposure}. As an alternative, ultrasound has a high sensitivity for detecting invasive cancer in dense breasts. Disadvantages of hand-held ultrasound(HHUS) in screening are: operator dependency, a long examination time, and the inability to acquire and archive 3-D volumetric images of the breast. Automated ultrasound approaches are developed to alleviate these problems\cite{tan2012computer}. Despite HHUS, Automated 3-D breast ultrasound (ABUS) screening process time is short and images are reproducible so they can be used for temporal comparison.

In ABUS imaging, for each breast, depending on its size, usually one to four 3-D views are acquired with different compression position on the breast. By Considering that up to 8 views are acquired for each patient, reading ABUS images is tedious, time-consuming, and subtle abnormalities may be missed \cite{tan2012detection}. Designing an accurate segmentation algorithm is important because it plays a main role in CADe(computer-aided detection systems that are used to find locations of abnormalities) and CADx(computer-aided diagnosis systems that are used to classify types of diseases) systems performance, which helps doctors to detect and diagnose breast tumors more accurately. Many researches \cite{tan2012computer}, \cite{tan2013computer}, \cite{tan2015computer} and \cite{kozegar2017breast} tried to develop CADe and CADx systems for 3-D ABUS images. Tan et al. \cite{tan2013evaluation}, also, evaluated the effect of computer-aided classification of benign and malignant lesions on reader performance in automated three-dimensional breast ultrasound. Moreover, Mass segmentation is a hard task for 3-D ABUS images because breast masses have a wide variety in shape, size and texture. Consequently, implementing a robust method that is invariant to these factors is challenging.

To the extent of our knowledge, three methods have been proposed for ABUS image segmentation. The first method was proposed by Kou et al.\cite{kuo2013automatic}. They used a radial gradient index(RGI) for initial contour determination. Afterward, they used the active model proposed by Li et. al.\cite{li2005level} for fine segmentation. This method was tested on a dataset that includes 64 masses from 55 patients. It achieved $0.60\pm0.17$, $0.57\pm0.18$ and $0.58\pm0.17$ overlap ratio on coronal, transversal and sagittal views respectively. In 2016 Tan et. al. \cite{tan2016segmentation} adopted a segmentation method that proposed by wang et al. \cite{wang2007segmentation} for lung nodule segmentation in CT images. In their segmentation method, they proposed a new criterion that states: if the voxel under consideration gets farther from the mass center, the probability of that voxel being part of the mass decreases. This method has the following steps: (1) spiral scanning to construct 2-D images from 3-D images (2) use dynamic programming to determine edges (3) applying multiple scanning directions to the model (4) reconstructing 3-D images and (5) using depth information to improve the results. Their method was applied on a dataset of 78 masses and achieved a mean Dice of $0.73\pm0.14$. Recently, Kozegar et al. \cite{kozegarMass} proposed a three-step method for mass segmentation. The steps are as follow: (1) noise reduction using Optimized Bayesian Non-Local Means (OBNLM) filter (2) pre-segmentation using a new region growing algorithm to obtain a rough estimation of a mass lesion (3) fine segmentation Using a modified version of Distance Regularized Level Set Evolution (DRLSE). Their method achieved a mean Dice of $0.74\pm 0.19$.

Many machine learning approaches can be used for different medical image analysis tasks like SVM, Random forest, logistic regression, etc. Besides, there is a newly emerged method, namely, deep learning. Deep learning is a novel machine learning approach that utilizes a range of neural network architectures to perform several machine learning tasks. The main difference between deep learning and conventional machine learning methods is in the feature extraction process. Conventional machine learning approaches use heuristic hand-crafted or engineered features with shallow feature learning architectures. However, deep learning approaches try to extract features automatically through a deep architecture.

In the last few years, deep convolutional neural networks outperformed other approaches in many visual recognition tasks like classification and segmentation. In classification tasks which is The typical use of these networks, a single label is assigned to the input image, while in segmentation tasks a class label is supposed to be assigned to each pixel of the input image. Most segmentation approaches have been developed for 2-D images while for 3-D ABUS images and other 3-D medical data new 3-D segmentation techniques are of interest.

Here, we introduce some segmentation techniques based on CNNs. Ciresan et. al.\cite{ciresan2012deep} trained a network for predicting membrane pixels in electron microscopy (EM) images in a sliding window, set up by providing a patch around the pixel as input. Indeed, they used a convolutional neural network (CNN) as a pixel classifier and an image is segmented by classifying all of its pixels. This method won the EM segmentation challenge at ISBI 2012. The proposed method has two drawbacks. First, since the network must be fed by each pixel, it is quite slow. Also, there is a lot of redundancy because of overlapping patches in adjacent pixels. Second, there is a trade-off between localization accuracy and use of context, i.e., as the patches get smaller, the network sees less context. On the other hand, as the patches become larger, more max-pooling layers are required that reduce the localization accuracy. Ronneberger et. al.\cite{ronneberger2015u} proposed a new architecture called U-net, which consists of a contracting encoder part and an expanding decoder part. The former part analyzes the whole input image and the latter part produces a full resolution segmentation image. While contracting path is composed of convolution and pooling layers, expanding path is composed of deconvolution and unpooling layers. U-net was a significant improvement in comparison with the classic CNN which was used by Ciresan et al. \cite{ciresan2012deep}. However, it was not able to process 3-D volumes. Cicek et. al.\cite{cciccek20163d} proposed a 3-D architecture based on U-net in which 2-D operators (e.g., 2-D pooling and 2-D convolution) were replaced with 3-D ones.

To the best of our knowledge, none of the CNN based mass segmentation methods have been used for 3-D ABUS images. In this paper, a new segmentation method based on U-net\cite{ronneberger2015u} is introduced for segmentation of 3-D ABUS images.

The rest of the paper is organized as follows. The proposed methods and dataset are introduced in section 2. Results obtained by the proposed methods are presented in section 3. Finally, conclusions are drawn in section 4.

\section{Materials and methods}
\subsection{Dataset}
\begin{figure}
	\centering
	\includegraphics[width=0.5\linewidth]{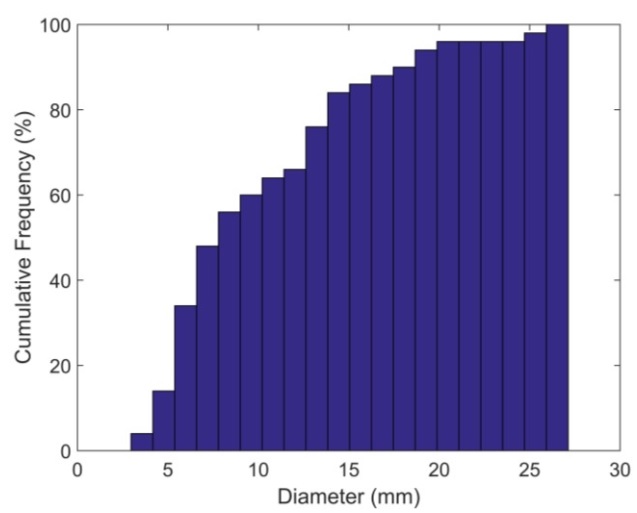}
	\caption{Cumulative histogram of lesion diameters of cancers in the dataset}
	\label{fig1}
\end{figure}

The dataset that is used in this study, was generated by two types of ABUS systems: the SomoVu automated 3-D breast ultrasound system developed by U-systems (Sunnyvale, CA, USA) and the ACUSON S2000 automated breast volume scanning system developed by Siemens (Erlangen, Germany). Images from SomoVu have a maximum size of $14.6cm X 16.8cm$ on the coronal plane and a maximum depth of 4.86 cm while images from ACUSON S2000 have a maximum size of $15.4cm X 16.8cm$ on the coronal plane and a maximum depth of 6 cm. The transducer of the U-systems device works with a fixed frequency of 8 or 10 MHz while the frequency of the transducer by Siemens is variable between 5 and 14 MHz, and is adjusted according to the breast size. Each 3-D volumetric view by SomoVu was generated with a minimum voxel size of 0.29 mm along the transducer, by 0.13 mm in the depth direction and by 0.6 mm along the sweeping direction while the images produced by ACUSON S2000 have a minimum voxel size of $0.21mm X 0.07mm X 0.5 mm$. To make the processing easier and faster, images were down-sampled to volumes with an isotropic voxel size of 0.6mm.
The presented dataset consists of 42 images (35 acquired by ACUSON  S2000 and 7 produced by SomoVu) from 32 patients; including 50 masses (38 malignant and 12 benign lesions). The mean diameter of the annotated masses in the dataset is $10.23 mm \pm 5.65 mm$. The cumulative sum of mass diameters is shown in figure \ref{fig1}. About 80\% of masses have diameters lower than 15 mm, and approximately 50\% of them have diameters smaller than 7 mm.

In a voxel-wise perspective, our dataset is entirely unbalanced. It has 25 Million non-lesion voxels and only 250,000 lesion voxels. Feeding the whole image to the network is not reasonable because the network will see too many non-lesion voxels, which is detrimental for the learning process. Cropping a patch around the mass center, omits a vast number of non-lesion voxels. Table \ref{table2} shows the impact of patch size on data balancing in our dataset. Although introducing a patch around the lesion center alleviates the imbalanced data problem, small patches cannot cover large lesions.

Our solution to the aforementioned problem is to crop patches around lesion centers and use them as a new dataset. The patch size is fixed and should be smaller than the size of the biggest lesion. This is done for the following reasons: (1) even the largest mass will fit in the patch using image scaling, (2) lesion to non-lesion voxels ratio will increase and network can learn with a more balanced dataset. We elaborate on our method in the next section.

\begin{table}
	\centering
	\caption{Lesion and non-lesion voxels for different patch sizes}
	\label{table2}
	\begin{tabular}{lllll}
		\hline
		patch size  & lesion voxels & non-lesion voxels & ratio &  \\ \hline
		300X300X100 & 272978        & 245584916         & 0.001 &  \\
		80X80X32    & 269457        & 9728879           & 0.028 &  \\
		40X40X32    & 195880        & 1081880           & 0.181 &  \\
		32X32X16    & 175912        & 642264            & 0.274 &  \\ \hline
	\end{tabular}
\end{table}


\subsection{The Proposed 3-D Segmentation Method}
The proposed algorithm has three main stages: preprocessing, segmentation using deep CNN and post-processing. We explain each stage in the next section.
\subsubsection{Pre-Processing}

Each modality has it is own difficulties which should be overcome via proper solutions. Ultrasound images suffer from speckle noise that decreases the quality of the images. Bayesian Non-Local Mean(OBNLM) filter \cite{coupe2009nonlocal} is used to reduce the impact of speckle noise.
\subsubsection{Convolutional Neural Network}

\begin{figure}
	\centering
	\includegraphics[width=0.9\linewidth]{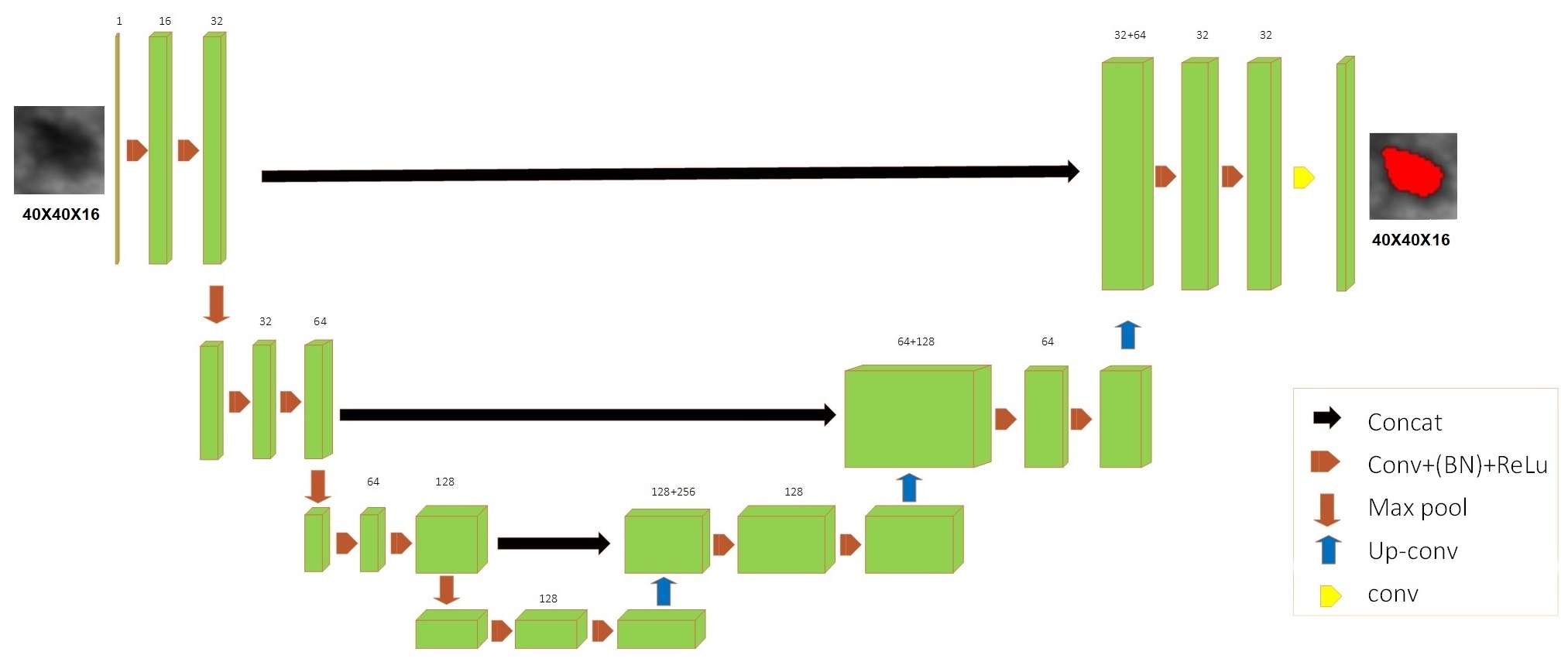}
	\caption{3-D U-net architecture. Green boxes represent feature maps. The arrows denote different operations}
	\label{fig2}
\end{figure}

The proposed network is similar to the U-net\cite{ronneberger2015u} and comprises a compression part and a decompression part. In the compression part, each layer contains two 3x3x3 convolutions each followed by a batch normalization (BN) and a ReLu and then a 2x2x2 max pooling with strides of two. In the decompression part, each layer consists of an up-convolution of 2x2x2 by strides of two, followed by two 3x3x3 convolutions each followed by a ReLu. Skip connections from layers of a similar resolution in the compression part provide extra features to the decompression part. In the last layer a 1x1x1 convolution reduces the number of output channels to the number of labels. The network is shown in figure \ref{fig2}.

\subsubsection{Post-Processing}

As we mentioned in the previous section, the network is fed with patches, with a size smaller than the biggest lesion, around the lesion center, but these patches cannot cover some large lesions in our dataset. Our solution for this problem is to consider a few down-sampling scales and For each case in the test phase do the following:

(1) feed the patch with specified size around the lesion center to the network and fetch output label map. (2) Count the number of positive labels that are in the patch boundaries and consider it as n. (3) If n is equal to zero, then return the label map as the final segmentation result. (4) If n is not zero then down-sample the image with next down-sampling scale and go back to step 2.
\subsection{The Proposed 2-D Segmentation Method}
Because many slices in a volume do not contain lesion voxels, another approach for reducing non-lesion voxels is to omit such slices from volumes in the dataset. To do so we assume that the indices of slices without lesions are the prior knowledge of the system. Having this knowledge, slices without lesions can be detected and omitted from the dataset. But the problem is that the new volumes will have a different number of slices, while the size of network input should be constant. For example, consider a volume that contains a big mass that spreads in twenty slices, and a volume that includes a small mass that spreads in only five slices. After omitting the non-lesion slices, The first volume size along the Z-axis will be 20 but the Second volume will have only 5 slices. The proposed solution is to extract 2-D images from 3-D volumes and use a 2-D network instead of a 3-D one.
Consequently, we can feed 3-D volumes to a 2-D network as 2-D images that have lesion voxels. A 2-D U-net with a similar configuration of Figure \ref{fig2} was used for this method. There is no need for post-processing because we consider a patch size of 80X80 and therefore, even the largest mass will be completely fitted in the patch. We have reported accuracy of 2-D network with the proposed post-processing method to make sure that the post-processing does not improve the performance of the network in the 2-D setting.

\begin{figure}
	\centering
	\includegraphics[width=0.9\linewidth]{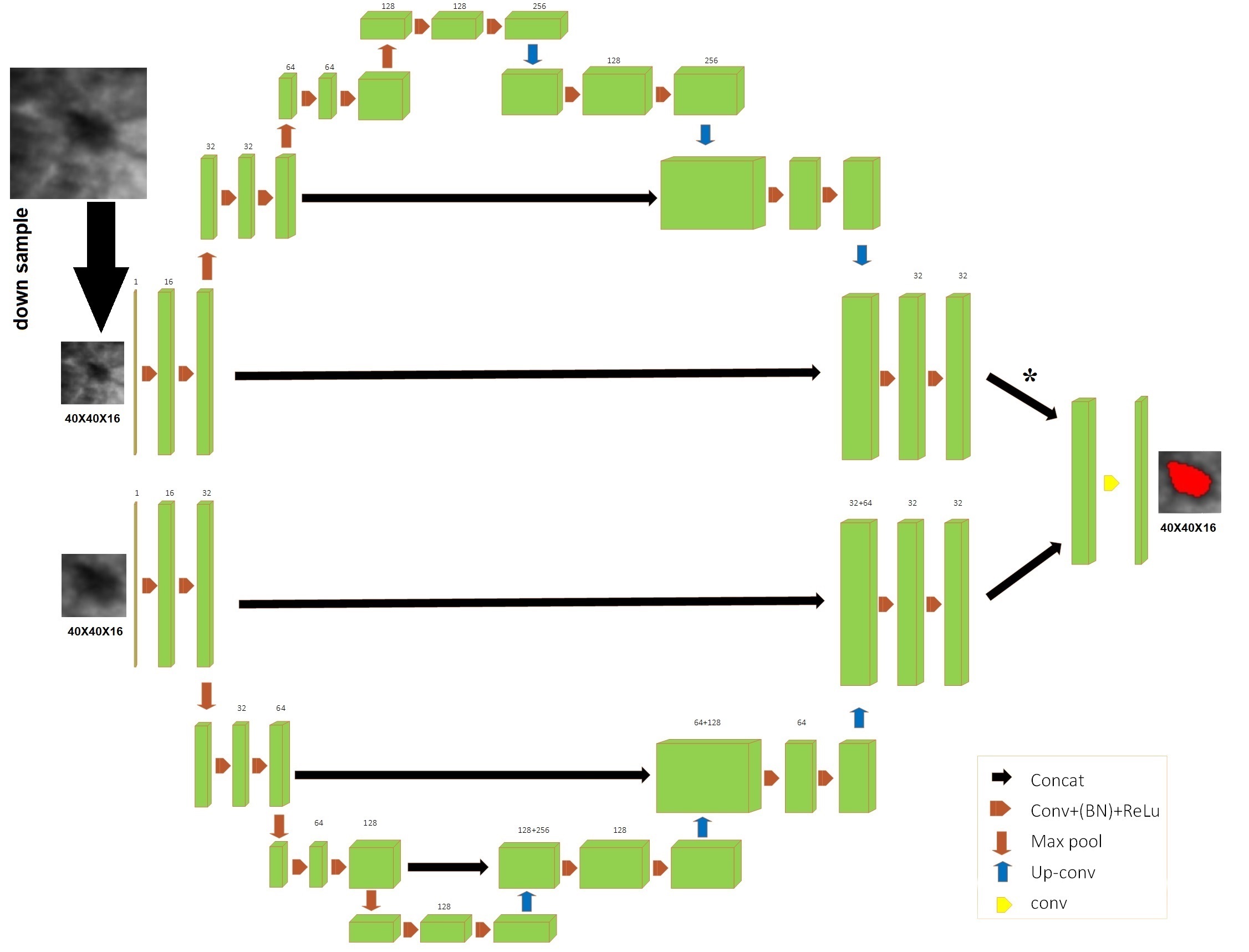}
	\label{fig22}
	\caption{Dual path U-net architecture. Green boxes represent feature maps. The arrows denote different operations}
\end{figure}

\subsection{The proposed Dual-path U-net}
Ghafoorian et al. \cite{ghafoorian2016non} proposed a multi-scale approach for white matter hyper intensity segmentation using deep convolutional neural networks. This approach can be used in U-net and for ABUS mass segmentation. Based on this idea we proposed a network that consists of two paths, each of them is an independent U-net. The first path is fed with volumes in original resolution and the second path is fed with same size volumes which are down-sampled to give a better overview of the volume to the network. Like U-net each path comprises a compression part and a decompression part. In the compression part, each layer contains two 3x3x3 convolutions each followed by a batch normalization (BN) and a ReLu and then a 2x2x2 max pooling with strides of two. In the decompression part, each layer consists of an up-convolution of 2x2x2 by strides of two, followed by two 3x3x3 convolutions each followed by a ReLu. Skip connections from layers of the similar resolution in the compression part provide extra features to the decompression part. In the last layer after concatenating the outputs of the two paths, a 1x1x1 convolution reduces the number of output channels to the number of labels. The proposed network is shown in \ref{fig22}.

\section{Experimental setup and Results}	
Algorithms and networks presented in this paper were implemented using python and Keras \cite{chollet2015keras} framework with TensorFlow \cite{abadi2016tensorflow} back-end on Windows 10. All the training and experiments were conducted on a standard workstation equipped with 16 GB of memory, an Intel(R) Core(TM) i7 CPU, and a Nvidia GTX 1070 with 8 GB of video memory. In the experiments, five-fold cross-validation strategy was used to evaluate the performance of the model. We used the Adam optimizer \cite{kingma2014adam} for optimization of the network variables and set all parameters to their default values.

The accuracy of the segmentation methods is measured in term of Dice Similarity Coefficient which measures the overlap between ground truth and the computerized segmented areas. This index is defined as

$$DSC\left(\mathrm{Im}1,\mathrm{Im}2\right)=\frac{2\left|im1\cap im2\right|}{\left|im1\right|+\left|im2\right|},$$
Where im1 and im2 are binary ground truth and binary segmented image, $|x|$ stands for the number of voxels of a given segmentation x, and $\cap$ is the intersection.

To demonstrate the fact that the proposed post-processing method has significantly improved the performance of the 3-D segmentation method, a system without post-processing was tested. Since all masses should be fitted in the patch with the selected size, omitting the post-processing stage limits the window size to sizes bigger than 80X80X32, which is bigger than the largest mass size in the dataset. Tables \ref{mytabone}, \ref{mytabtwo}, \ref{mytabthree} summarize the performance of 3-D U-net, 2-D U-net, and comparison with other methods, respectively.

\begin{table}
	\centering
	\caption{Performance of the proposed 3-D methods with different downsampling scales}
	\label{mytabone}
	\begin{tabular}{llll}
		\hline
		Patch size & path & Down-sampling Scales         & DSC \\ \hline
		80x80x16   & single             & without post processing      & 0.59    \\
		40x40x16   & single             & 0.9, 0.8, 0.7, 0.6, 0.5      & 0.76    \\
		40x40x16   & single             & 0.75, 0.5                    &  0.74   \\
		32x32x16   & single             & 0.9, 0.8, 0.7, 0.6, 0.5, 0.4 &  0.76   \\
		32x32x16   & single             & 0.8, 0.6, 0.4                &  0.75   \\
		40x40x16   & double             & 0.9, 0.8, 0.7, 0.6, 0.5      &  0.77   \\
		40x40x16   & double             & 0.75, 0.5                    &  0.76   \\
		32x32x16   & double             & 0.9, 0.8, 0.7, 0.6, 0.5, 0.4 &  0.76   \\
		32x32x16   & double             & 0.8, 0.6, 0.4                &  0.75   \\
		32x32x8   & double             & 0.9, 0.8, 0.7, 0.6, 0.5&  0.75   \\
            &              & 0.4, 0.3, 0.2 &    \\
		32x32x8   & double             & 0.8, 0.6, 0.4, 0.2                &  0.74   \\ \hline
	\end{tabular}
\end{table}

\begin{table}
	\centering
	\caption{Performance of proposed 2-D method with different downsampling scales}
	\label{mytabtwo}
	\begin{tabular}{llll}
		\hline
		Patch size & path & Down-sampling Scales      & DSC \\ \hline
		80x80   & single             & WO postprocessing     & 0.82    \\
		40x40   & single             & 0.9, 0.8, 0.7, 0.6, 0.5      & 0.74    \\
		40x40   & single             & 0.75, 0.5                    & 0.74    \\
		32x32   & single             & 0.9,0.8,0.7& 0.73    \\
           &              & 0.6,0.4 &   \\
		32x32   & single             & 0.8, 0.6, 0.4                & 0.72    \\ \hline
	\end{tabular}
\end{table}

\begin{table}
	\centering
	\caption{Performance comparison of the proposed methods with other works}
	\label{mytabthree}
	\begin{tabular}{ll}
		\hline
		Method   & DSC \\ \hline
		Spiral scanning\cite{tan2016segmentation} & 0.73   \\
		Adaptive Region Growing and Edge-Based Deformable Model\cite{kozegar2017mass} & 0.74   \\
		3-D U-net without post-processing & 0.59   \\
		Single-path 3-D U-net & 0.76   \\
		Dual-path 3-D U-net & 0.77   \\
		2-D U-net &  0.82 \\ \hline
	\end{tabular}
\end{table}

Mass shape complexity is one of the main challenges of mass segmentation algorithms. In this paper, the compactness measure has been used to determine the complexity of masses shapes. It is calculated as:
$$compactness = \frac{{A}^{3}}{36\pi {\nu }^{2}}$$
Where A is the enclosing surface of a 3-D mass and V is its volume. Figure \ref{fig:compactness} plots the segmentation accuracy (DSC) against the shape complexity (compactness) for test samples. As indicated in the figure, masses with complex shapes also achieved a remarkable accuracy. 

Another feature that can affect the accuracy of segmentation algorithms is the size of masses. We had a large variety of masses with different sizes in our dataset. To show that how the size of masses affects the accuracy of the proposed 3-D method, relation of accuracy and size of the masses are shown in figure \ref{fig:precisonSize}. 

In addition to the mentioned criteria, the proposed 3-D method can be examined to find the relation between predicted mass compactness and ground truth compactness per mass. In figure \ref{fig:gt-pred-comp} the compactness of ground truth and the predicted mass for each sample are shown. As can be seen, most of the predicted boundaries for masses are simpler than what they really are.
\begin{figure}
	\centering
	\includegraphics[width=0.8\linewidth]{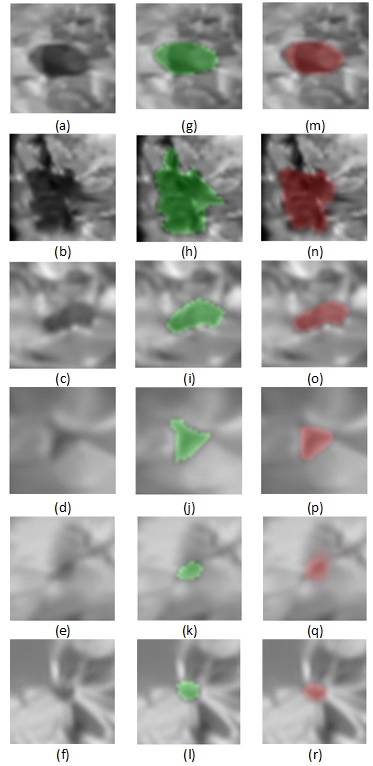}
	\caption{An example of the results of the proposed 3D method; (a)-(f): six consequent slices of a breast volumetric region of interest, (g)-(l): green regions are the corresponding regions delineated by an expert, (m)-(r): red regions are the outputs of the proposed segmentation method.}
	\label{fig:example}
\end{figure}

\begin{figure}
	\centering
	\includegraphics[width=0.9\linewidth]{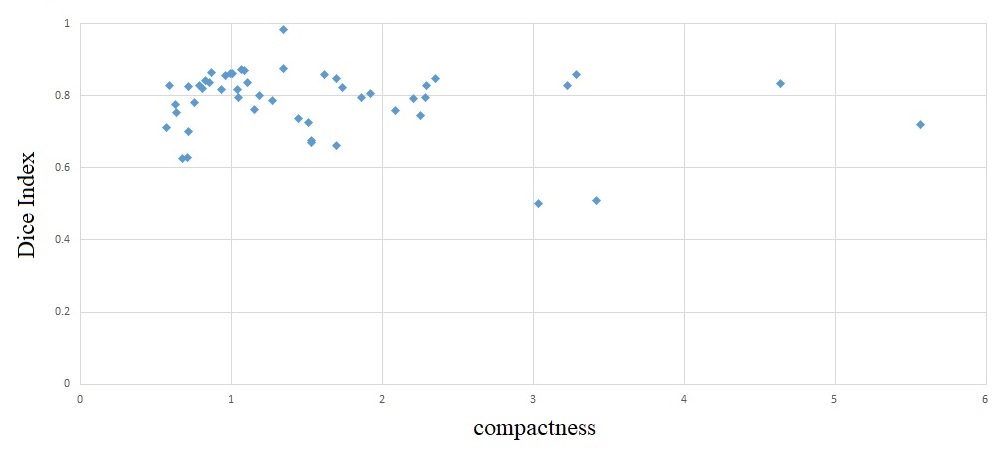}
	\caption{Evaluation of the robustness of the algorithm against mass complexity by illustrating the relation between DSC of each sample and its ground truth compactness}
	\label{fig:compactness}
\end{figure}
\begin{figure}
	\centering

	\includegraphics[width=0.9\linewidth]{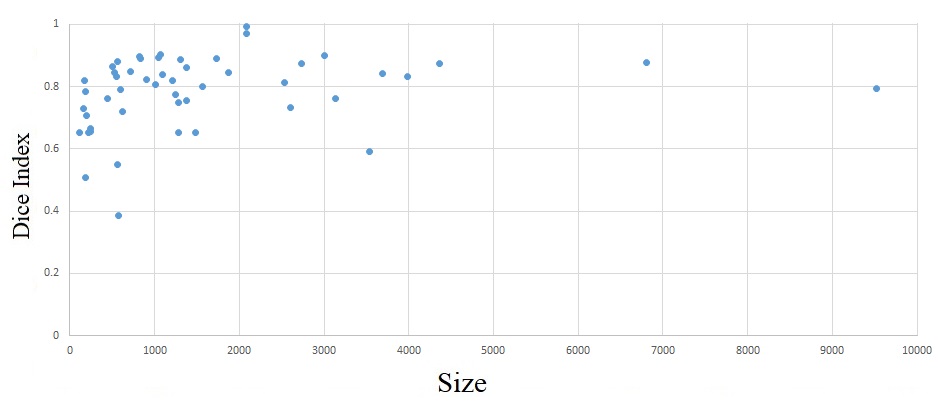}
	\label{fig:precisonSize}
	\caption{Evaluation of the robustness of the algorithm against mass size by illustrating the relation between DSC of each sample and its ground truth size}
\end{figure}

\begin{figure}
	\centering
	\includegraphics[width=0.9\linewidth]{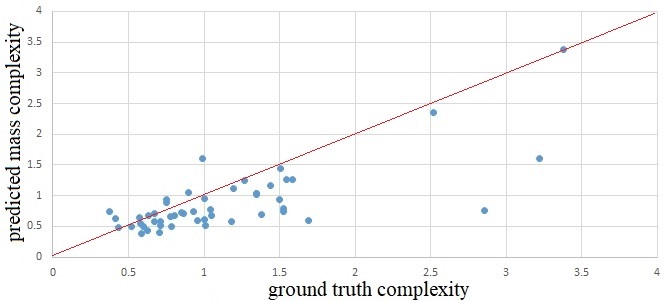}
	\label{fig:gt-pred-comp}
	\caption{The relation between each sample compactness and its predicted region compactness}
\end{figure}

An example result of the proposed method is shown in figure \ref{fig:example}

\section{Conclusions and discussions}
Segmentation of masses in 3-D ABUS images is a challenging task due to the large variety in mass size, shape, and intensity. Furthermore, lesion to non-lesion voxel ratio is too high in these kind of datasets. Consequently, there are not enough lesion voxels to train a complicated network. 

Prior to our work, three researches had been done for 3D ABUS images and the methods and results were published in \cite{kozegar2017mass}, \cite{tan2016segmentation} and \cite{kuo2013automatic}. They achieved mean dice values of 0.74, 0.73, and 0.70 respectively. None of these methods used deep learning as their approach. 

In this paper, we presented novel approaches based on deep learning and the best method achieved a mean dice of 0.82. In other words, our methods outperform the three previous methods by using a deep learning approach that inspired from a well-known deep network architecture called U-net. The presented work introduced three different methods for mass segmentation in 3-D ABUS images based on deep learning: (1) a basic method that uses U-net with a constant patch size (as large as the biggest mass size). (2) A more advanced 3-D method that reduced the size of the patch to diminish the unbalancity problem of the dataset and added a post-processing stage to make sure all large masses would be fitted in the patch by recursive down-sampling. (3) A 2-D method that feeds a 2-D U-net, only with slices that have lesion voxels.

Each method can have up to 3 stages: (1) preprocessing for denoising, (2) CNN for segmentation, and (3) post-processing to make sure that the mass is not larger than the patch which is cropped from the original volume. This stage is added just when patch size is smaller than the largest mass size.

A disadvantage of these methods is their result boundary smoothness. Because many features that are extracted from masses after segmentation have a high dependency on the attributes of their boundaries, inaccurate and smooth edges can have detrimental effects on the performance of other components (e.g. mass classification) that use segmentation results.

Our dataset only contains 50 cases, which is a limiting factor of this study. Furthermore, one radiologist have annotated our dataset masses. If more radiologists annotated our dataset masses, we could obtain more accurate ground truth. So, gathering larger and more accurate datasets can open up new opportunities for training more complex and advanced models in the future.

\bibliography{U-net-Hamed}

\end{document}